\documentclass[12pt]{article}
\usepackage[utf8]{inputenc}
\usepackage[english]{babel}
\usepackage{amssymb,graphicx}
\pdfoutput=1
\usepackage{graphics}
\usepackage{amsmath,amsfonts}
\usepackage{fullpage}
\usepackage{wrapfig} 
\usepackage{braket}
\usepackage[colorinlistoftodos]{todonotes}

\renewcommand{\tanh}{\mathop{\rm th}\nolimits}

\graphicspath{{images/}{figures/}}

\usepackage{import}
\usepackage{xifthen}
\usepackage{pdfpages}
\usepackage{transparent}

\usepackage{xcolor}

\title{
\begin{flushright}
{\normalsize INR-TH-2025-009}
\end{flushright}
\vspace{0.5cm}
Entanglement islands and black hole decay \\ in regular dilaton gravity}
\author{Maxim Fitkevich$^{1,2}$ \vspace{.2cm}\\
\normalsize\it $^1$ Institute for Nuclear Research of the Russian Academy
of Sciences, \\  
\normalsize \it  60th October Anniversary Prospect, 7a, 117312  Moscow, Russia\\
\normalsize\it $^2$ Moscow Institute of Physics and Technology, \\  
\normalsize \it Institutskiy per., 9, 141701 Dolgoprudny, Moscow Region, Russia\\  
\normalsize \it fitkevich@phystech.edu
}

\begin{document}

\maketitle

\abstract{We consider a class of two-dimensional dilaton gravity models with linear dilaton vacuum including Callan--Giddings--Harvey--Strominger (CGHS) model as the special case. General thermodynamic properties of black holes in such models are evaluated. We focus on the CGHS model and its modification with regular black holes as empty--space solutions characterized by ever--present finite curvature. We find generalized entanglement entropy blows--up for near--extremal regular black holes considered as remnants. That signalling a possible breakdown of the semiclassical approximation near the endpoint of evaporation. We conjecture that remnants are unstable and decay by quantum fluctuations into horizonless spacetimes. We give an estimate for the decay amplitude by using a semiclassical regularization method and propose a path to mitigate the unitarity loss problem.}

%%%%%%%%%%%%%%%%%%%%%%%%%%%%%%%%
\section{Introduction}\label{sec:intro}
Unitarity of quantum gravity became a puzzling topic after Hawking revealed quantum black holes do evaporate into almost thermal radiation~\cite{Hawking:1975vcx}. It seemed very reliable to conclude from basic QFT on curved spacetime one has a final quantum state described by maximally entangled density matrix at the very endpoint of evaporation. Therefore, it necessarily implies that non--unitary evolution is unavoidable in processes involving black holes~\cite{Hawking:1976ra}. At first glance, the gravitational interaction can not be incorporated into quantum theory before drastic modifications of the latter.

Replacing well--established quantum postulates is logically possible but very problematic from theoretical standpoint. A league of researches in opposition to initial Hawking's claim followed the conservative route which is to find a self--consistent theory of quantum gravity where unitary evolution is automatically secured~\cite{Harlow:2014yka}. It was successfully fulfilled in holographic models including ones stemmed from the AdS/CFT correspondence~\cite{Ryu:2006bv}. One of significant recent advancements in this field is a derivation of quantum extremal surfaces from ``replica wormholes''~\cite{Penington:2019npb,Almheiri:2019qdq} in various theories of gravity including ones that possess asymptotically flat spacetime solutions~\cite{Hartman:2020swn,Gautason:2020tmk}.

It was found that extremal surfaces directly relate the entanglement entropy of quantum fields with geometry what became codified in a form of so--called {\it island rule} which allowing to derive a unitary Page curve for the Hawking radiation entropy~\cite{Page:1993wv}. Nevertheless, a complete understanding of dynamical mechanisms behind recovery of quantum information from black holes is still elusive. For instance, we lack derivation of the unitary S--matrix describing processes of gravitational scattering~\cite{Stephens:1993an}.

One can argue that the S--matrix elements between quantum states of collapsing matter and Hawking radiation can be calculated semiclassically at least in principle. A dominant channel for black hole evaporation is a decay into large number of soft particles which validates evaluation of the path integral,
\begin{equation}\label{eq:pi}
\langle f|\hat{{\cal S}}|i\rangle =\int_{\Phi(t_i)=\Phi_i}^{\Phi(t_f)=\Phi_f}{\cal D}\Phi\,e^{\frac{i}{\hbar}I[\Phi]}\approx F[\Phi_0]e^{\frac{i}{\hbar}I[\Phi_0]}(1+{\displaystyle O}(\hbar))\;, \quad\quad \frac{\delta I}{\delta\Phi}[\Phi_0]=0\;,
\end{equation}
through generally complex--valued saddle points $\Phi_0$ solving the classical equations of motion with boundary conditions $\Phi_{i,f}$ corresponding to given in-- and out--states. 

The problem one encounters is strongly coupled dynamics near singularity which necessarily breaks the semiclassical approximation. One may try to bypass bad singularities in ``complex plane'' by saddle points deformation. Regularization methods applied to simplified model yield physically meaningful answers consistent with expectations from unitarity~\cite{Bezrukov:2015ufa,Fitkevich:2020tcj}.

Another useful approach is to implement the limiting curvature condition,
\begin{equation}\label{eq:limit-curv}
R_{\mu\nu\sigma\rho}R^{\mu\nu\sigma\rho}\leq l_\ast^{-4}\;,
\end{equation}
to avoid singularities on the level of classical field equations~\cite{Markov,Markov:1984ii}. Evidently, one can not realise this kind of behaviour without violation of energy conditions. In this scenario a vicinity of black hole singularity on a scale $r\simeq l_\ast$ can be turned either into de Sitter core or bounced back~\cite{Bardeen,Simpson:2018tsi}. Such regular black holes possess global event horizons which can be dissolved by quantum backreaction~\cite{Hayward:2005gi,Frolov:2014jva}. Non--singular black holes appear in string theory and non--commutative geometry~\cite{Nicolini:2023hub}, loop quantum gravity~\cite{Ashtekar:2020ifw}, and other modified theories of gravity~\cite{Chamseddine:2019pux}.

We continue these lines of research in our paper concerning some toy models of two--dimensional dilaton gravity which are supposed to be the best choice for testing new ideas about quantum properties of black holes~\cite{Grumiller:2002nm,Almheiri:2014cka,Narayan:2020pyj,Trunin:2020vwy}. Such system can be potentially even modelled on quantum computers~\cite{Pikulin:2017mhj}. It is widely assumed that transverse degrees of freedom can be integrated out near horizon so that gravitational scattering is described by effectively two-dimensional models \cite{Verlinde:1991iu,Gukov:2022oed}. These ones capture many relevant features but lack certain annoying problems of its higher--dimensional ancestors including infamous non--renormalizability of the Einsteinian GR~\cite{Giddings:1992ae}.

Peculiar properties of two--dimensional quantum theories allow us to tune these models to satisfy Eq.~\eqref{eq:limit-curv} on the level of classical field equations. In this paper we continue our research of two--dimensional dilaton gravity model with non--singular (regular) black holes~\cite{Fitkevich:2022ior} (for similar attempts see Refs.~\cite{Easson:2002tg,Frolov:2021kcv}).

The proposed regular model has a parameter $M_\mathrm{ext}$ which corresponds to the extremal black hole mass. Being the lightest the extremal black holes are considered stable because of zero temperature so it can potentially be a sink for information. All matter content that fallen into black hole remains inside and the entanglement entropy of Hawking radiation can grow indefinitely. Therefore, this state can be considered as a {\it black hole remnant} with infinite number of internal states~\cite{Aharonov:1987tp}. It is usually assumed that remnants violate the Bekenstein bound and can not be present in unitary theories but there are objections to this claim~\cite{Marolf:2008tx,Rovelli:2017mzl}.

We show by calculation of the entanglement entropy of Hawking radiation via the island formula that aforementioned concerns about remnants may not be realised. What we conjecture instead is that the extremal state tends to decay at early stages. Backreaction makes black holes to tunnel and we estimate a quantum transition amplitude by finding saddle--point solutions for null shock waves penetrating the event horizon.

Instead of solving a genuine quantum field theory we simplify the model so that the shock wave is represented a single massless particle. Therefore, a system of gravity equations radically simplifies and one has two--parametric solution describing transitions from spacetime with mass $M_i$ to spacetime with mass $M_f$. For $M_i,\,M_f<M_\mathrm{ext}$ the classical solution is real so one evaluates Eq.~\eqref{eq:pi} in a straightforward manner. However, if final or initial states contain black holes one need to perform analytic continuation to obtain saddle--point solutions. It is shown that probability of transition with black holes has exponential suppression,
\begin{equation}
{\cal P}_{fi}=|\langle f|\hat{{\cal S}}|i\rangle|^2\simeq\exp(-\Delta S_\mathrm{BH})\;,
\end{equation}
where $\Delta S_\mathrm{BH}$ is a difference of black hole entropies before and after the null shock wave released from under horizon.

The obtained result has an interpretation which is consistent with unitary expectation as following. The number of black hole states is proportional to $\exp(S_\mathrm{BH})$ so that probability of emitting a hard particle in a final state is expected with probability of order $\exp(-S_\mathrm{BH})$. Therefore, entropic suppression takes place for an exclusive transition process. One expects a total probability ${\cal P}_{fi}\simeq 1$ for all possible decay modes. The most probable decay channel is emitting of a Hawking radiation particle with energy of order $T_\mathrm{H}$ if mass is large. Black hole decays completely if its mass is comparable with $M_\mathrm{ext}$.

The paper is organized as follows. In Section~\ref{sec:models} we introduce a general action for linear dilaton gravity and find empty--space solutions of the field equations, and describe black hole thermodynamics. In Section~\ref{sec:islands} we calculate the entanglement entropy of the Hawking radiation and show it blows up near the end of evaporation for non--singular model. In Section~\ref{sec:decay} we discuss regular black holes near the endpoint of evaporation and conclude they are unstable. We discuss our results and prospects in Section~\ref{sec:discussion}. We leave technical details of calculations to Appendices.

%%%%%%%%%%%%%%%%%%%%%%%%%%%%%%%%
\section{Dilaton gravity toy models}\label{sec:models}
We consider a subclass of general dilaton models described by the gravitational action,
\begin{equation}\label{eq:grav-action}
I_\mathrm{ldg}=\int d^2x\sqrt{-g}\left(W(\phi)R+W''(\phi)\left[(\nabla\phi)^2+\lambda^2\right]\right)
\end{equation}
where the dilaton field $\phi$ non--minimally couples to the Ricci scalar $R$ with function $W(\phi)$ specifying choice of model and primes denote derivatives with respect to $\phi$. The dimensionful parameter $\lambda$ sets the characteristic energy/length scale in the model.

The general solution with no matter is
\begin{equation}\label{eq:swd-solution}
\phi=-\lambda r\;,\qquad ds^2=-f(r)dt^2+\frac{dr^2}{f(r)}\;, \qquad f(r)=1+\frac{M}{\lambda W'(\phi)}\;,
\end{equation}
where integration constant $M$ can be identified with the Arnowitt--Deser--Misner (ADM) mass for asymptotically flat spacetimes with function $f(r)\to 1$ as $r\to\infty$. We derive field equations following from Eq.~\eqref{eq:grav-action} in Appendix~\ref{app:field-eqs}. The Minkowski spacetime ($M=0$) is a solution usually called the {\it linear dilaton} vacuum which gave its name to models described by Eq.~\eqref{eq:grav-action}. Uniqueness of the empty space solution is guaranteed by the two--dimensional analogue of the Birkhoff theorem~\cite{Louis-Martinez:1993agn}. A set of points $r=r_\mathrm{hor}$ where $f(r_\mathrm{hor})=0$ corresponds to event horizons or Cauchy horizons. 

Thermodynamic properties of the black holes are derived from Euclidean solutions of field equations given by the Wick rotation $t\mapsto \tau=it$ in Eq.~\eqref{eq:swd-solution}. These are periodic ``cigar'' solutions which are smooth everywhere except for maybe a single conifold singularity at $r=r_\mathrm{hor}$ which is removed if one equates the imaginary time period with the inverse Hawking temperature $\beta_\mathrm{H}={T_\mathrm{H}}^{-1}$,
\begin{equation}\label{eq:haw-t}
T_\mathrm{H}=\frac1{4\pi}\frac{df}{dr}(r_\mathrm{hor})=\frac{\lambda^2W''(\phi_\mathrm{hor})}{4\pi M}\;,
\end{equation}
where we used Eq.~\eqref{eq:swd-solution} in the second equality. 

%Next, one defines the partition function,
%\begin{equation}
%{\cal Z}(T_\mathrm{H})=\int{\cal D}\Phi\,e^{-I_\mathrm{E}[\Phi]}
%\end{equation}
%where $I_E$ is the euclidean action, first time calculated by Gibbons and Hawking as euclidean path integral. 

%From the standard formula, $S=\beta^2\partial_\beta I_\mathrm{E}-I_\mathrm{E}$, one straightforwardly finds the Bekenstein--Hawking entropy from Eq.~\eqref{eq:haw-t} for the linear dilaton gravity,
By using the Wald formula one finds the Bekenstein--Hawking entropy,
\begin{equation}\label{eq:bh-ent}
S_\mathrm{BH}=\Bigl.\frac{4\pi}{\sqrt{-g}}\frac{\delta I_\mathrm{ldv}}{\delta R}\Bigl|_{r=r_\mathrm{hor}}=4\pi W(\phi_\mathrm{hor})\;,
\end{equation}
where the horizon value of the dilaton field $\phi_\mathrm{hor}=-\lambda r_\mathrm{hor}$ serves as a substitute for invariant area of horizon which is just a single point in two dimensions.

From now we focus on two dilaton gravity models. The first one with $W(\phi)=e^{-2\phi}$ is the prominent CGHS model~\cite{Callan:1992rs,Russo:1992ht,Russo:1992ax} which have interpretation as a spherical reduction of the 1+3-dimensional Horndeski theory \cite{Mandal:2023kpu}. The second one is the regular black hole (sinh--CGHS) model defined by $W(\phi)=-(M_\mathrm{ext}/2\lambda)\sinh(2\phi)$ where the classical singularity is avoided.

For empty--space solutions in the sinh--CGHS model one has
\begin{equation}\label{eq:et-reg-bh}
f(r)=1-\frac{M}{M_\mathrm{ext}\cosh(2\lambda r)}\;,
\end{equation}
which describes regular black holes if $M\geq M_\mathrm{ext}$. The outer and inner horizons are situated at $r=r_\mathrm{hor}$, 
\begin{equation}\label{eq:sh-cghs-hor}
r_\mathrm{hor}=\frac1{2\lambda}\log\left(\frac{M}{M_\mathrm{ext}}+\sqrt{\left(\frac{M}{M_\mathrm{ext}}\right)^2-1}\right)\;,
\end{equation}
and $r=-r_\mathrm{hor}$ correspondingly. 

A solution with $M=M_\mathrm{ext}$ is the extremal black hole, and solutions with $M<M_\mathrm{ext}$ are the horizonless spacetimes. Note that arbitrary negative values of the mass parameter $M$ are not prohibited by anything in the classical theory. The latter can sway the stability of the model in presence of matter fields what we discuss afterwards.

A geodesically complete spacetime which can be locally described by Eq.~\eqref{eq:et-reg-bh} is an infinite tower of eternal black and white holes interconnected by its interiors~\cite{Fitkevich:2022ior}. It seems reasonable to consider such spacetimes with interconnected asymptotically flat regions as non-physical in dynamical situations including gravitational collapse and evaporation. One may wonder if it is possible that such solutions contribute to Eq.~\eqref{eq:pi} without spoiling unitarity of quantum gravity. We assume it is possible to define the gravitational path integral so that such configurations with eternal black holes are effectively excluded. With that in mind we simply ignore these solutions in current discussion.

From Eqs.~\eqref{eq:haw-t}, \eqref{eq:bh-ent}, and \eqref{eq:et-reg-bh} one derives the Hawking temperature,
\begin{equation}\label{eq:sh-cghs-T}
T_\mathrm{H}=\frac{\lambda}{2\pi}\sqrt{1-\left(\frac{M_\mathrm{ext}}{M}\right)^2}\;,
\end{equation}
and the Bekenstein--Hawking entropy in the sinh--CGHS model,
\begin{equation}\label{eq:sh-cghs-S}
S_\mathrm{BH}=\frac{2\pi}{\lambda}M\sqrt{1-\left(\frac{M_\mathrm{ext}}{M}\right)^2}\;,
\end{equation}
which reduce to the CGHS model in the limit $M_\mathrm{ext}\to 0$. Note that both the temperature and the entropy vanish for the extremal ($M=M_\mathrm{ext}$) black holes. 

Formally, the model~\eqref{eq:grav-action} enters the strong coupling regime as $W(\phi)\to 0$ what requires additional considerations. We expect it could probably be not a problem because propagating gravitons are absent at the one--loop level. The reason is that dilaton gravity degrees of freedom are highly constrained; in two dimensions the Einstein--Hilbert action itself is nothing more than topological Euler characteristic. Indeed, one can make an arbitrary shift $W(\phi)\mapsto W(\phi)+\mathrm{const}$ which does not affect field equations but moves a supposed strong coupling zone anywhere we want. We analysed sinh--CGHS static solutions with quantum conformal anomaly and came to conclusion the apparent strong coupling is in fact irrelevant for singularity avoidance~\cite{Fitkevich:2022ior}. To summarize, the backreaction is proportional to $R$ which is finite and solutions become asymmetrically tilted but still everywhere regular. We postpone a thorough analysis of dynamical solutions in presence of quantum matter to next publications.

%%%%%%%%%%%%%%%%%%%%%%%%%%%%%%%%
\section{Entropy from entanglement islands}\label{sec:islands}
In this section we consider quantum fields on the eternal black hole background. First, we calculate the entanglement entropy of the Hawking radiation for trivial setup and reproduce a version of the information paradox for eternal black holes. Second, we apply the island formula prescription to obtain corrected expression for the entanglement entropy in the CGHS model. Third, we repeat this calculation for the regular black holes and compare a difference between results.

\begin{figure}[h]
\centerline{
\hspace{0.2cm}\includegraphics[width=12.18cm]{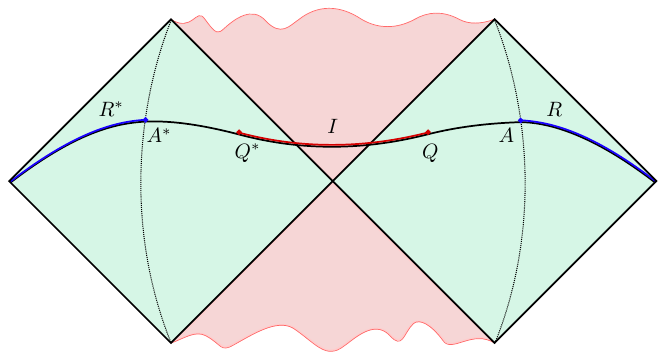}
}
%\vspace{-4mm} \hspace{3.5cm}(a) \hspace{7.5cm}(b)
\caption{ Penrose diagram depicting setup for entanglement island. Legend: A, A$^*$ are anchor points, R, R$^*$ are parts of Cauchy surface containing old Hawking radiation, Q, Q$^*$ are QEPs, and I is island. Wavy line simbolizes geodesic incompleteness of spacetime either with singularities or Cauchy horizons. } \label{fig:island-setup}
\end{figure}

We consider a conformal matter with large central charge ${c}$ to make sense of the semiclassical approximation where the spacetime is a real smooth manifold. For such a thing one makes a nice slicing of the whole spacetime labelled by some global time coordinate. Each slice is a Cauchy surface divided into regions of interest as depicted in Fig.~\ref{fig:island-setup}.

Asymptotic regions R and R$^*$ extend from anchor points A and A$^*$ correspondingly to spacelike infinities and contain quantum radiation far from black hole. One may think of observers situated at A and A$^*$ as ones who collect Hawking quanta passing through them in dynamical situation. For the black hole not to evaporate it should stay at equilibrium with the outside thermal bath. Therefore, the quantum fields are described by the Hartle--Hawking state $\hat{\rho}_\mathrm{HH}$ for which the energy--momentum tensor is regular across the event horizon.

The quantum fields living on the union of R and R$^*$ are represented by the reduced density matrix,
\begin{equation}
\hat{\rho}_R=\underset{AA^*}{\mathrm{Tr}}(\hat{\rho}_\mathrm{HH})\;,
\end{equation}
where the partial trace is taken over the states living on the interval AA$^*$ containing the Einstein--Rosen bridge.

In what follows we evaluate the von Neumann entropy of the outside radiation,
\begin{equation}
S_\mathrm{ent}[R\cup R^*]=-\mathrm{Tr}\;(\hat{\rho}_R\log\,\hat{\rho}_R)=S_\mathrm{ent}[AA^*]\;,
\end{equation}
where the second equality follows directly from purity of the whole quantum state.

The entanglement entropy for conformal fields with central charge ${c}$ can be calculated explicitly~\cite{Holzhey:1994we}. For modes living on the compact spacelike curve AA$^*$
\begin{equation}\label{eq:von-N-ent}
S_\mathrm{ent}[AA^*]=\frac{c}{6}\log\left(\epsilon^{-2}(v_A-v_{A^*})(u_{A^*}-u_A)\right)+\frac{c}{6}(\rho_A+\rho_{A^*})\;,
\end{equation}
where $\epsilon$ is an ultraviolet cutoff and factor $\rho$ comes from the metric in conformally flat coordinates,
\begin{equation}\label{eq:conf-metr}
ds^2=-e^{2\rho(v,u)}dvdu\;,
\end{equation}
so that the proper length of the cutoff remains coordinate independent.

One introduces a coordinate transformation,
\begin{equation}\label{eq:conf-trans}
v=\frac1{2\pi T}e^{2\pi T(t+\bar{r})}\;, \qquad\qquad u=-\frac1{2\pi T}e^{-2\pi T(t-\bar{r})}\;,
\end{equation}
where the ``tortoise'' coordinate
\begin{equation}\label{eq:tortoise}
\bar{r}(r)=\int^r\frac{dr'}{f(r')}\;,
\end{equation}
to relate Eqs.~\eqref{eq:swd-solution} and \eqref{eq:conf-metr} by formulas,
\begin{equation}\label{eq:rho}
\rho(v,u)=\frac12\log(f(r)/g(r))\;, \qquad\qquad g(r):=\exp(4\pi T\bar{r}(r))\;,
\end{equation}
for the eternal black hole solutions.

Without loss of generality one assumes the anchor points $A$ and $A^*$ are symmetric with respect to the spatial inversions so that $u_{A^*}=v_A$, $v_{A^*}=u_A$, and $\rho_A=\rho_{A^*}$ in the conformally flat coordinates. Substituting Eqs.~\eqref{eq:conf-trans} and \eqref{eq:rho} into Eq.~\eqref{eq:von-N-ent} one finds the entanglement entropy,
\begin{equation}\label{eq:haw-ans-ent}
S_\mathrm{ent}[R\cup\bar{R}]=\frac{c}{3}\log(2\cosh(2\pi T t_A))+\frac{c}{6}\log(f(r_A))-\frac{c}{3}\log(2\pi T\epsilon)\;,
\end{equation}
which aligns with expected linear growth from the Hawking result $\dot{S}\simeq 2\pi c\,T/3$ at large times $t_A\to +\infty$.

Evidently, the naive semiclassical answer~\eqref{eq:haw-ans-ent} violates the Bekenstein bound for the black hole with surrounding thermal radiation already having maximal entropy,
\begin{equation}
S_\mathrm{ent}[R\cup R^*]\leq 2(S_\mathrm{BH}(M)+S_\mathrm{rad})\;,
\end{equation}
which implies saturation of the entanglement entropy instead of unbounded growth to allow unitary evolution. This is a version of the information paradox for eternal black holes.

It was conjectured that Eq.~\eqref{eq:von-N-ent} for the Hawking radiation entropy gives correct answer for early times only. If backreaction is included unexpected dominant contributions to the gravitational path integral with non-trivial topologies may arise at late times~\cite{Stanford:2022fdt}. It was first discovered in the AdS/CFT context but extended as a conjecture onto any consistent quantum gravitational model. Direct calculations in various gravity models supported this claim~\cite{Hartman:2020swn}.

Idea is that gravity influences quantum fields and one should add so--called an {\it entanglement island}, see region I in Fig.~\ref{fig:island-setup}. The island is an additional region of spacetime representing nonperturbative gravitational contributions in the exact path integral. Usual von Neumann entropy is then replaced with generalized entropy functional defined for the eternal black hole as follows,
\begin{equation}\label{eq:gen-ent}
S_\mathrm{gen}[R,I]=S_{\mathrm{grav}}[\partial I]+S_{\mathrm{ent}}[R\cup I\cup R^*]\;,
\end{equation}
where the first term is the Bekenstein contribution from pure gravity and the second term comes from von Neumann entropy of quantum fields living on the union of $R$ and $R^*$, and the island $I$.

%\begin{figure}[ht]
%    \centering
%    \resizebox{.8\linewidth}{!}{\incfig{island-calculation}}
%    \caption{Penrose diagram depicting setup for entanglement island.}
%    \label{fig:island-calculation}
%\end{figure}

The island is bound by the quantum extremal surface which are points Q and Q$^*$ in 1+1 dimensions. The recipe for calculating a true value of the entanglement entropy is given by a formula,
\begin{equation}\label{eq:receipt}
S_\mathrm{ent}[R\cup R^*]=\underset{I\in\{\varnothing,\,QQ^*\}}{\mathrm{min}}\;\underset{\partial I}{\mathrm{ext}}\,S_\mathrm{gen}[R,I]\;,
\end{equation}
where one have to vary positions of the quantum extremal points (QEPs) first and then take a minimal value of the generalized entropy among solutions including one with island and one without it leading to Eq.~\eqref{eq:haw-ans-ent}.

For large spacelike separations of $Q$ and $Q^*$ the approximation $S_{\mathrm{ent}}[R\cup I\cup R^*]\simeq S_{\mathrm{ent}}[QA]+S_{\mathrm{ent}}[Q^*A^*]$ becomes valid. Without loss of generality we choose anchor points and QEPs to be symmetric on both sides of the eternal black hole. With all given assumptions the generalized entropy, 
\begin{equation}\label{eq:gen-app}
S_\mathrm{gen}(x_Q,x_A,M)=8\pi W(\phi_Q)+\frac{c}{3}\log(\epsilon^{-2}(v_A-v_Q)(u_Q-u_A))+\frac{c}{3}(\rho_A+\rho_Q)\;,
\end{equation}
is just twice the contribution from a single side of the black hole. 

The first extremality condition $\partial S_\mathrm{gen}/\partial t_Q=0$ gives $t_Q=t_A$. Substitution into Eq.~\eqref{eq:gen-app} removes time dependence,
\begin{multline}\label{eq:gen-formula-no-time}
S_\mathrm{gen}(r_Q,r_A,M)=8\pi W(-\lambda r_Q)+\frac{2c}{3}\log\left(\sqrt{g(r_A)}-\sqrt{g(r_Q)}\right)+\frac{c}{6}\log\left(\frac{f(r_Q)f(r_A)}{g(r_Q)g(r_A)}\right) \\
-\frac{2c}{3}\log(2\pi T\epsilon)\;,
\end{multline}
where we expressed everything in terms of Schwarzschild coordinates.

The second extremality condition $\partial S_\mathrm{gen}/\partial r_Q=0$ yields a transcendental equation relating $r_Q$ with black hole mass parameter $M$. We find $r_Q(M)$ for the CGHS and sinh--CGHS models in App.~\ref{app:islands}. 

Finally, one calculates the entanglement entropy,
\begin{equation}\label{eq:reg-ent}
S_\mathrm{ent}[R\cup R^*]=S_\mathrm{gen}(r_Q(M),r_A,M)+S_0(c,\epsilon,r_A)\;, \qquad r_A\to +\infty\;,
\end{equation}
but fixing the infrared regulator $S_0$ to remove divergence in Eq.~\eqref{eq:gen-formula-no-time} and shift the entropy so it is sensible zero in flat spacetime limit $M\to 0$. 

\begin{figure}[t]
\centerline{
\hspace{0.2cm}\includegraphics[width=9.93cm]{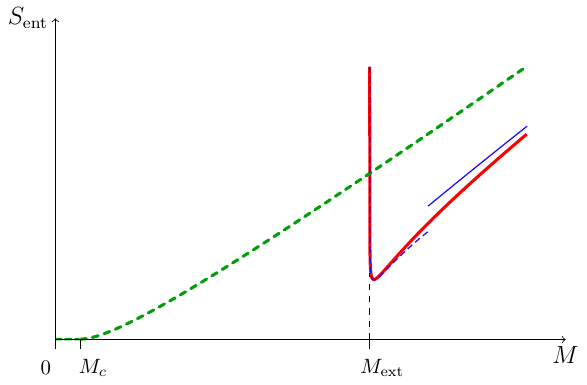}
}
%\vspace{-4mm} \hspace{3.5cm}(a) \hspace{7.5cm}(b)
\caption{ Plots of generalized entropy of the Hawking radiation for the CGHS model \eqref{eq:ent-cghs} (dashed) and the sinh-CGHS model \eqref{eq:exact-sh-cghs-gen} (thick). The asymptotic behaviour is given by Eq.~\eqref{eq:large-m} at $M\to\infty$ (thin) and Eq.~\eqref{eq:sh-ent-mext} at $M\to M_\mathrm{ext}$ (dashed thin). } \label{fig:island-ent}
\end{figure}

In case of the CGHS model one finds the entropy,
\begin{equation}\label{eq:ent-cghs}
S_\mathrm{ent}[R]=\left[\frac{4\pi}{\lambda}M-\frac{c}{6}\log\left(\frac{M}{2\lambda}\right)+\frac{c}{6}\left(\log\left(\frac{c}{48\pi}\right)-1\right)\right]\theta(M-M_c)\;,
\end{equation}
which goes to zero at $M\leq M_c=c\lambda/24\pi$, see in Fig.~\ref{fig:island-ent}. Interestingly, the quantum--corrected entropy from Ref.~\cite{Solodukhin:1995te} and Eq.~\eqref{eq:ent-cghs} are similar.

In case of the sinh--CGHS model one finds the entanglement entropy given by Eq.~\eqref{eq:exact-sh-cghs-gen} which diverges as $M\to M_\mathrm{ext}$, see in Fig.~\ref{fig:island-ent}. It prevents us from setting $S_\mathrm{ent}(M_\mathrm{ext})$ to zero. What we do instead is setting renormalization term in Eq.~\eqref{eq:reg-ent} so that it asymptotically matches Eq.~\eqref{eq:ent-cghs} at large masses $M\to +\infty$,
\begin{equation}\label{eq:large-m}
S\simeq\frac{4\pi}{\lambda}M-\frac{2\pi}{\lambda}\frac{{M_\mathrm{ext}}^2}{M}-\frac{c}{6}\log\left(\frac{M}{2\lambda}\right)+\frac{c}{6}\left(\log\left(\frac{c}{48\pi}\right)-1\right)+{\displaystyle O}\left(\frac{{M_\mathrm{ext}}^2}{M^2}\right)\;,
\end{equation}
which can be interpreted as large black holes in the CGHS model and regular black holes should not be distinguished in experiments performed by distant observer.

With all given assumptions one finds from expansion of Eq.~\eqref{eq:exact-sh-cghs-gen} at $M\simeq M_\mathrm{ext}$ a near--extremal behaviour for the entanglement entropy,
\begin{equation}\label{eq:sh-ent-mext}
S\simeq 4\pi\sqrt{\frac{2M_\mathrm{ext}(M-M_\mathrm{ext})}{\lambda^2}}+\frac{c}{6}\left(\log\left(\frac{2M_c}{M-M_\mathrm{ext}}\right)-1\right)-\frac{c}{3\sqrt{2}}\sqrt{
\frac{M}{M_\mathrm{ext}}-1}\;,
\end{equation}
where aforementioned divergence manifests. It drastically differs from the naive expectation given by Eq.~\eqref{eq:sh-cghs-S}.

One can see that for near--extremal black holes the information paradox stays apparently unresolved. The Bekenstein--Hawking entropy goes to zero and the entanglement entropy eventually exceeds the finite thermodynamic entropy of radiation. The formula \eqref{eq:receipt} reads one has to switch back to Hawking answer \eqref{eq:haw-ans-ent} which is non--unitary.

This drawback of the island formula is not attributed specifically to the sinh--CGHS model considered above. It was shown the the usual GR black holes suffer from the same problem in the zero mass limit~\cite{Arefeva:2021kfx}. Moreover, even additional boundary condition far from black hole itself can spoil unitarity of the final answer~\cite{Ageev:2023hxe}. 

Therefore, one may argue that the island prescription is not an ultimate solution to the information paradox and something yet has to be answered about the endpoint of the Hawking evaporation. We turn to this issue in the next Section.

%%%%%%%%%%%%%%%%%%%%%%%%%%%%%%%%
\section{Quantum decay of remnants}\label{sec:decay}
%\subsection{Semiclassical picture of evaporation}
In this Section we analyse a whole process of the regular black hole evaporation. Ignoring backreaction we consider process as a sequence of quasistatic black holes with fading mass according to a time dependence $M(t)$. The latter follows directly from the 1+1--dimensional Stefan–Boltzmann law,
\begin{equation}
\dot{M}=-\frac{\pi c}{12}{T_H}^2\;,
\end{equation}
which has an exact solution for the sinh--CGHS black holes,
\begin{equation}\label{eq:evap-law}
M(t)+\frac{M_\mathrm{ext}}{2}\log\left(\frac{M(t)-M_\mathrm{ext}}{M(t)+M_\mathrm{ext}}\right)=\mathrm{const}-\frac{\lambda^2 c}{48\pi}t\;,
\end{equation}
where the integration constant is related to the initial black hole mass. From Eq.~\eqref{eq:evap-law} one can see that the regular black holes approach the extremal state only asymptotically considering the semiclassical approximation is valid at late times. Therefore, the extremal black holes are stable and can be considered as remnants\footnote{There were signs that quantum remnants appear in the one--loop CGHS model~\cite{Almheiri:2013wka}.}.

\begin{figure}[t]
\centerline{
\hspace{0.2cm}\includegraphics[width=11.1cm]{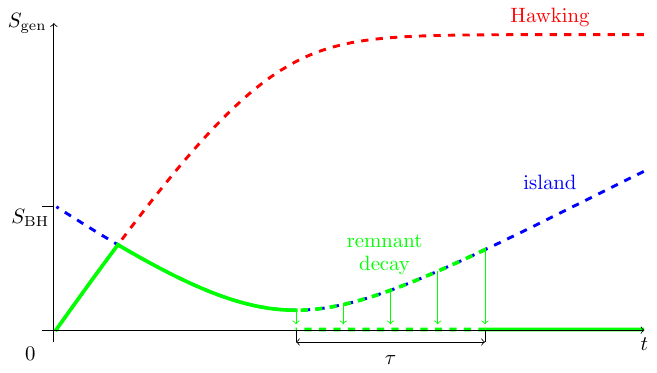}
}
%\vspace{-4mm} \hspace{3.5cm}(a) \hspace{7.5cm}(b)
\caption{ Page curve for sinh--CGHS black hole during evaporation. Monotonic dashed curve is Hawking answer for radiation $S_\mathrm{rad}(t)=2(S_\mathrm{BH}(0)-S_\mathrm{BH}(t))$. Contribution from island falls before it approaches $M_\mathrm{qb}$ and then starts to grow again exceeding Hawking answer. Remnants should decay to protect unitary evolution on timescale $\tau$. } \label{fig:page-curve}
\end{figure}

The entanglement island calculation from the previous Section suggests that black hole follows the unitary Page curve right before it reaches quantum breaking point as it demonstrated in Fig.~\ref{fig:page-curve}. After that the entanglement entropy grows again until it reaches the Bekenstein entropy of initial black hole what clearly violates unitarity.

In old scenarios remnants contained a large amount of entanglement and minimal possible energy. Wilczek specifically stressed that purification can happen at almost zero energy cost \cite{Wilczek:1993jn}. Indeed, unlike thermodynamic one the entanglement entropy is not defined locally and can not be attributed to piece of matter for good. In order to purify the Hawking radiation remnants have to decay into huge number of soft particles which implies a horrendous half-life but without any dynamical mechanism explaining its stability \cite{Carlitz:1986nh}.

One can turn this conclusion around to say that it actually means for remnants to decay at the beginning into few quanta where the Page curve reaches its minimum. The whole process of gravitational scattering should not be merely treated as combination of classical collapse with subsequent thermal evaporation. One needs to apply quantum--mechanical rules to understand full dynamics behind the endpoint of black hole evolution. The fact that Eq.~\eqref{eq:evap-law} implies infinite lifetime for remnants means non-equilibrium dynamics becomes important.

%\subsection{Deviations from thermality}
We do expect deviations from Eq.~\eqref{eq:evap-law} at large times. It can be argued that thermodynamic description is not applicable in the near-extremal case \cite{Preskill:1991tb}. The consistency condition for semiclassical approximation is a requirement that emission of single Hawking quantum shall not change the black hole temperature significantly so that
$\Delta T\simeq T/C$ where $C$ is a large number. A typical quantum reduces black hole mass by amount $\Delta M\simeq T$ assuming thermality of the Hawking radiation \cite{Visser:2014ypa}. One derives a condition
\begin{equation}\label{eq:consistency}
\frac{\Delta M}{\Delta T}\simeq\frac{dM}{dT}=C\gg 1\;,
\end{equation}
where the large number $C$ is identified with the black hole heat capacity. Recalling from fluctuation theory $\langle(\delta T)^2\rangle=T^2/C$ and $\langle(\delta S)^2\rangle=C$ one realises that relative temperature fluctuations become large and entropy fluctuations become small while approaching the extremal state as $C\to 0$.

The quantum breaking of semiclassical approximation corresponds to violation of condition~\eqref{eq:consistency}. It happens for black holes with mass less than
\begin{equation}\label{eq:th-break-mass}
M_\mathrm{qb}\simeq M_\mathrm{ext}+\frac{\lambda^2}{8\pi^2M_\mathrm{ext}}\;,
\end{equation}
which possess only a few degrees of freedom because the entropy $S\lesssim{\displaystyle O}(1)$ in such cases. The controlling semiclassical parameter $M_\mathrm{ext}/\lambda\gg 1$ in Eq.~\eqref{eq:th-break-mass} is the same as standing in front of the classical action~\eqref{eq:grav-action}. Statistical description is applicable at zero temperature for systems represented by an infinite number of degrees of freedom what is far from being the case for the near--extremal black holes of mass $M\lesssim M_\mathrm{qb}$ and temperature $T\lesssim\lambda^2/M_\mathrm{ext}$.

%It is known result that extremal black holes in 3+1--dimensions possess non--zero entropy proportional to horizon area but still vanishing temperature. Therefore, the extremal black holes are represented by a degenerate quantum state. One may ask if those degrees of freedom decouple in some way allowing to set $S\;\mapsto\;S-S_\mathrm{ext}$ or can be probed in gravitational scattering off the extremal black hole. We do not focus on this issue here assuming $S_\mathrm{ext}\equiv 0$ in the dilaton gravity.

Notice that each emission of Hawking quantum drops the black hole entropy by $\Delta S=(dS/dM)\Delta M\simeq(dS/dM)T=1$, where we used the first law of thermodynamics in the last equality. Under such observations the Hawking radiation near the endpoint resembles more the shot noise than continuous energy flux \cite{Gray:2015pma}. In this case it is more appropriate to imagine black hole evaporation as sequential two--body decays.

One may expect that remnants decay with a significant probability as an average Hawking quantum energy reaches a mass gap left, $M_\mathrm{qb}-M_\mathrm{ext}\simeq E\simeq T_H(M_\mathrm{qb})$. From Eqs.~\eqref{eq:evap-law} and \eqref{eq:th-break-mass} one evaluates the corresponding time scale,
\begin{equation}\label{eq:lifetime}
\tau=48\pi\frac{M_\mathrm{ext}}{c\lambda^2}\left(\log\left(4\pi\frac{M_\mathrm{ext}}{\lambda}\right)+{\displaystyle O}\left(\frac{\lambda^2}{M_\mathrm{ext}^2}\right)\right)\;,
\end{equation}
after which only ${\displaystyle O}(1)$ emissions are left for the remnant to decay. After emitting the last drop of energy the ADM mass can become lower than $M_\mathrm{ext}$ and the event horizon dissolves completely leaving empty space.

%%%%%%%%%%%%%%%%%%%%%%%%%%%%%%%%
%\section{Black sprays as candidate saddles}\label{sec:sprays}
%\begin{quote}
%{\small {\it That’s it, the black sprays. That’s a good name. Well, you know their properties. If you shine a ray of light into one of those beads, the transmission of the light is delayed and the delay depends on the bead’s weight, size, and several other parameters. And the unit of light coming out is always smaller than the one entering. What is this? Why? There is a wild theory that the black sprays are gigantic expanses of space with properties different from those of our space and that they became curled up under the influence of our space.}} \cite{Strugatsky}
%\end{quote}

We construct a spacetime corresponding to the decaying black hole in the sinh-CGHS model. First, we introduce the Vaidya solution,
\begin{equation}\label{eq:vaidya}
ds^2=-\left(1-\frac{{\cal M}(v)}{M_\mathrm{ext}\cosh(2\lambda r)}\right)dv^2+2dvdr\;,
\end{equation}
where ${\cal M}(v)$ is the Bondi mass, for derivation see Appendix~\ref{app:field-eqs}. The ingoing and outgoing null rays follow from $ds=0$ and Eq.~\eqref{eq:vaidya}.

\begin{figure}[t]
\centerline{
\hspace{0.2cm}\includegraphics[width=12.7cm]{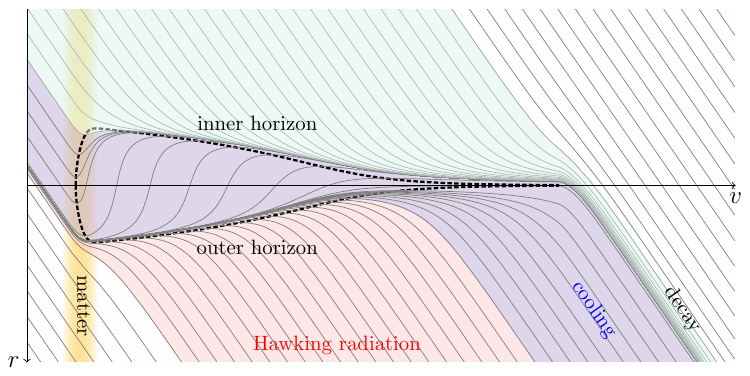}
}
%\vspace{-4mm} \hspace{3.5cm}(a) \hspace{7.5cm}(b)
\caption{ Spacetime diagram for solution of Eq.~\eqref{eq:vaidya}. Thin lines are null rays satisfying equation $M_\mathrm{ext}\cosh(2\lambda r(v))(1-2r'(v))={\cal M}(v)$. Vertical strip represents wave packet of matter forming initial black hole with closed apparent horizon (dashed curve). Black hole evaporates linearly producing a bulk of Hawking radiation. On the final stage it cools down forming a remnant which eventually decays into last burst of quantum radiation leaving empty space. } \label{fig:black-spray}
\end{figure}

We mimic the anticipated behaviour as following\footnote{Such decaying solutions were numerically found in a similar model of the dilaton gravity with the sprerically reduced Bardeen black hole as a solution~\cite{Barenboim:2024dko} (see also Hayward black holes in two dimensions~\cite{Cadoni:2023tse}). The one--loop correction was introduced as a quantum source causing the black hole to evaporate. It was found that inclusion of backreaction rendered regular black holes to be dynamically unstable in contrast to naive expectations from Eq.~\eqref{eq:evap-law}.}. First, the Bondi mass reaches the value $M_i$ as wave packet collapses into black hole. One can use shock wave approximation. Second, the black hole evaporates according to Eq.~\eqref{eq:evap-law} until it reaches quantum breaking scale $M_\mathrm{qb}$ triggering a decay modelled by hand. In the end the Bondi mass can be any $M_f<M_\mathrm{ext}$. The resulting spacetime is shown in Fig.~\ref{fig:black-spray}. %We call it the black spray.

%Instead of apparent horizon $F(v,r)=0$ we use quasi-horizon satisfying the equation
%\begin{equation}
%2\partial_vF(v,r)+F(v,r)\partial_rF(v,r)=0\;,
%\end{equation}
%which serves as an attractor both for outgoing null geodesics near presumable position of the inner horizon and for outgoing null geodesics near the outer horizon propagated in backward time direction \cite{Frolov:2014jva}.

%Light rays accumulates on the inner branch of the quasi-horizon and escape after the end of evaporation. We pointed out that sihn-CGHS model is not potentially safe from mass inflation as radiation could potentially accumulate close to the inner horizon. For the black spray the stress tensor of finite but it can disrupt the original metric and backreaction must be taken into account. One should treat this result very carefully since inclusion of backreaction on the metric can drastically change endpoint behaviour like finite relaxation time down to extremal state.

%It was argued that remnants are present in high-dimensional gravitational theories on the scale which is significantly larger than Planck's so its are weakly coupled \cite{Germani:2017ceg}. 

One needs to point out a feasible pathological behaviour. Since Eq.~\eqref{eq:et-reg-bh} makes sense at arbitrary negative mass parameter $M$ the matter fields can backscatter dynamically so to extract an unbounded amount of energy from the gravity--matter system. Therefore, the regular model can appear unstable exactly because of singularity absence. It was conjectured that classical singularities actually should persist so the negative mass solutions forced to be excluded from gravitational path integral to make it consistent on the quantum level~\cite{Horowitz:1995ta}. This latter issue can be cured by forbidding the matter fields to enter the second asymptotic region so it scatters back. The simplest choice is to introduce a running mass $\sim e^{\phi}$ of the matter fields or the reflecting dynamical boundary at $\phi=\phi_0$~\cite{Fitkevich:2017izc,Fitkevich:2020okl}.

%\subsection{Transition amplitude}
Let us estimate the black hole decay probability from purely quantum--mechanical viewpoint. In order to do so we consider the near--extremal black hole supported by a sharp wavepacket right under the inner horizon. The tunneling event disrupts the horizon allowing the wavepacket to get out of the interior. We assume it is reliable to model the wavepacket by a single massless particle\footnote{We need to stress that backreaction on the metric can be significant because of the mass--inflation problem~\cite{Poisson:1989zz}. Quantum corrections can dump the magnitude of this effect but not enough to treat it as a small perturbation.}. 

Given initial and final quantum states $\Psi_i$ and $\Psi_f$ the transition amplitude is represented by the path integral,
\begin{equation}\label{eq:path-int}
{\cal A}_{fi}:=\langle\Psi_f|\hat{{\cal S}}|\Psi_i\rangle=\lim_{\substack{t_f\to +\infty\\t_i\to -\infty}}\int {\cal D}\Phi\,e^{iI_0(0,t_f)+iI(t_f,t_i)+iI_0(t_i,0)}\Psi_f^\ast[\Phi]\Psi_i[\Phi]\;,
\end{equation}
where the integration measure includes matter, dilaton, metric, and ghost fields. We explicitly set the time contour ``encircling infinity'' in the total action including contributions from interacting and free time evolution operators in the formal definition of the scattering operator $\hat{{\cal S}}=\hat{{\cal U}}_0(0,+\infty)\cdot\hat{{\cal U}}(+\infty,-\infty)\cdot\hat{{\cal U}}_0(-\infty,0)$. It is important that boundary conditions for the fields $\Phi$ are not identical at different endpoints $t_{f,i}=0$ of the time contour.

We evaluate semiclassical amplitude using saddle--point approximation for Eq.~\eqref{eq:path-int},
\begin{equation}
{\cal A}_{fi}\simeq e^{i I_\mathrm{tot}[\Phi_\mathrm{cl}]}\;,
\end{equation}
where $I_\mathrm{tot}=I_0^i+I_0^f+I-i\log(\Psi_f^\ast\Psi_i)$ is the total action and $\Phi_\mathrm{cl}$ is generally complex--valued solution of field equation with boundary conditions corresponding to initial and final states.

Instead of the field theory we consider a massless particle with energy $E=M_i-M_f$ representing a shock wave passing through gravity region $|\phi|\lesssim 1$. Mass parameters $M_i$ and $M_f$ correspond to patches of spacetime before and after the shock wave moving from left to right.

The amplitude is a pure phase, for calculation see Appendix~\ref{app:ampl},
\begin{multline}
{\cal A}_{fi}(M_f,M_i)\simeq\exp(i\varphi(M_i)-i\varphi(M_f))\;, \\
\varphi(M)=\frac{M}{\lambda}-\frac{2M_\mathrm{ext}}{\lambda}\sqrt{1-\frac{M^2}{M_\mathrm{ext}^2}}\arctan\sqrt{\frac{M_\mathrm{ext}+M}{M_\mathrm{ext}-M}}\;,
\end{multline}
for real solution with $M_i,\,M_f<M_\mathrm{ext}$.

In order to obtain correct saddle--point solutions with $M_i,\,M_f\geq M_\mathrm{ext}$ one needs to perform an analytic continuation. We use a semiclassical method described in Appendix~\ref{app:regul}. One can show that correct amplitudes are restored as,
\begin{equation}\label{eq:anal-cont}
{\cal A}_{fi}(M_f,M_i)=\lim_{\varepsilon\to +0}{\cal A}_{fi}(M_f+i\varepsilon,M_i+i\varepsilon)\;,
\end{equation}
for regularized solutions having a slightly complex mass.

We find the transition probability is exponentially suppressed,
\begin{equation}\label{eq:probability}
{\cal P}_{if}(M_f,M_i)=|{\cal A}_{fi}|^2=\exp(-S_{BH}(M_i)+S_{BH}(M_f))\;,
\end{equation}
for the shock wave tunnelling through horizons of the black hole with mass $M_i$ and leaving the black hole with mass $M_f$ in final state.

Considering the shock--wave has undefined energy for external observers it is instructive to average over the final mass,
\begin{equation}\label{eq:averaging}
\langle M_f\rangle=\frac1{N_0}\int_{M_\mathrm{ext}}^{M_i}dM\,M\,{\cal P}_{if}(M,M_i)\;, \qquad\quad N_0=\int_{M_\mathrm{ext}}^{M_i}dM\,{\cal P}_{if}(M,M_i)\;,
\end{equation}
where Eq.~\eqref{eq:probability} is interpreted as a probability distribution for an ensemble of black holes with shock wave inside interior. By expanding Eq.~\eqref{eq:averaging} into Taylor series at large $M_i$ one finds,
\begin{equation}\label{eq:M-exp}
\langle M_f\rangle=M_i-T_\mathrm{H}(M_i)\;,
\end{equation}
which is in agreement with thermodynamic expectations. It can be demonstrated that Eq.~\eqref{eq:M-exp} stops working long before $M_\mathrm{qb}$ is reached. Numerical calculation shows it happens at $\Delta M={\displaystyle O}(M_\mathrm{ext})$.

%%%%%%%%%%%%%%%%%%%%%%%%%%%%%%%%%
\section{Discussion}\label{sec:discussion}
In this paper we compared the entanglement entropy of Hawking radiation for black holes in the CGHS model and for regular black holes in the sinh--CGHS model using the island formula. It was found out that the entropy follows the Page curve before asymptotically stable extremal state is formed in the regular model but blows up afterwards. Therefore, one concludes (meta)stable remnants violate unitarity and must decay fast in order to preserve quantum evolution.

We estimated the decay amplitude using semiclassical regularization method. We modelled the process by the massless pointlike particle travelling across the regular spacetime. Exact calculation yields the S--matrix element ${\cal A}_{fi}(M_f,M_i)$ depending on mass parameters on both sides of the particle worldline. If energies are low the amplitude is a pure phase and the transition occurs completely classically. At larger energies the amplitude becomes exponentially suppressed by a factor $\exp(-\Delta S_\mathrm{BH}/2)$ where $\Delta S_\mathrm{BH}=S_\mathrm{BH}(M_i)-S_\mathrm{BH}(M_f)$ is a change in black hole entropy before and after the transition process. It can be interpreted as a probability of emitting a single particle with energy $E=M_i-M_f$ by the black hole with mass $M_i$. Additional information inside the phase of the transition amplitude is not so easily interpreted.

This result brings us to a conclusion that extremal black holes are unstable and decay by quantum fluctuations without formation of long--living remnants. It was recently suggested that black hole remnants can constitute stable dark matter~\cite{Bianchi:2018mml,Rovelli:2018okm} but our finding contradicts to this hypothesis\footnote{It needs to be stressed that conclusions of our paper should not be taken for granted in higher dimensions which requires additional analysis because extremal black holes have nonzero Bekenstein entropy in general~\cite{Carroll:2009maa}.}. At first glance that means regular black holes model allows for unitary evolution: the whole spacetime becomes horizonless and there is nothing preventing matter to escape from would--be interior.

However, there is still a caveat. Imagine a situation where the black hole kept at approximately same mass by sending wave packets for a long time continuing indefinitely a process of information consumption. Quantum field theory on the curved spacetime should describe this situation satisfactory because for each Cauchy slice there is a quantum state on a smooth classical spacetime. But the island entropy calculation signifies that Hawking radiation must contain a second copy of the internal state. This gives a bound for quantum correlations to be transferred somehow across the apparent horizon before its dissolution.

One may consider the scrambling time~\cite{Hayden:2007cs},
\begin{equation}
t_\mathrm{scr}=\frac1{2\pi T_H}\log\,S_{BH}\;,
\end{equation}
as an estimate for time necessary for quantum information to escape from the interior. Evidently, one may prolong black hole existence so that evaporation time $t_\mathrm{evap}\gg t_\mathrm{scr}$. The regular spacetime has no room for quantum fields to leak out before the moment of decay what brings us to the xerox paradox similarly to a conclusion from analysis of the charged near--extremal black holes \cite{Strominger:1993yf}.

\begin{figure}[t]
\centerline{
\hspace{0.2cm}\includegraphics[width=8.40cm]{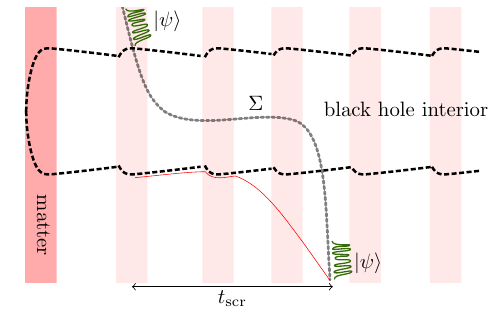}
}
%\vspace{-4mm} \hspace{3.5cm}(a) \hspace{7.5cm}(b)
\caption{ Quantum xerox. Vertical strips are wave packets falling into black hole. After scrambling time the quantum information returns with Hawking radiation quanta (thin line). One can make a slicing $\Sigma$ so that quantum state $|\psi\rangle$ is represented simultaneously inside the black hole interior and in the Hawking radiation. } \label{fig:xerox}
\end{figure}

Therefore, the semiclassical spacetime is not what should actually represent black hole evaporation process because it did not capture quantum information recovery at all\footnote{We bring attention to this observation that unitary evolution can be a problem even on the regular spacetimes without event horizons. It was advocated that area--law entropic bounds are universal for any consistent unitary quantum field model~\cite{Dvali:2020wqi}. Nevertheless, a precise meaning of the Bekenstein bound is still under debate~\cite{Hayden:2023ocd}.}. We conjecture some sort of tunnelling solutions similar to the real--time gravitational instantons. One may wonder if such saddle points of the Lorentzian path integral are related to the ``replica wormholes'' by complicated analytic continuation. The regularization method used to calculate transition amplitudes provides a consistent framework for searching these complex--valued solutions.

As a final remark we outline some directions for future developments. One proposes to test unitarity of the S--matrix numerically by evaluating the path integral in coherent states formalism~\cite{Tinyakov:1992dr},
\begin{equation}\label{eq:regularized-S}
\langle b|\hat{{\cal S}}|a\rangle=\int{\cal D}c_k^*{\cal D}c_k\;e^{-\int dk\;c_k^*c_k}\langle b|\hat{{\cal S}}_\epsilon|c\rangle\langle c|a\rangle\;,
\end{equation}
where $\hat{{\cal S}}_\varepsilon$ is the S-matrix acting on the subspace corresponding to horizonless spacetime boundary conditions in theory with genuine quantum fields.

If successful this regularization method applied to the toy regular model can be promoted to usual 3+1 dimensional gravity. Notably, the infamous non-renormalizability problem can be irrelevant to the information paradox. It was shown by resummation of diagrams with virtual gravitons exchanging momenta between in- and out-particles on the black hole background in certain limit of the effective field theory that non-perturbative answer is insensitive to UV physics \cite{Gaddam:2020mwe}.

We expect the relevant saddle points bypass singularities in the ``complex plane'' so that the strong coupling regime is avoided and semiclassical calculation of the gravitational path integral is justified. Interesting question if such solutions can be mapped onto the ``effective geometry'' reflecting dynamics behind the scattering process involving intermediate black holes and inheriting some properties of regular solutions. Effects may include lightcone tilting induced by spacetime fluctiations which allows for apparent FTL propagation near singularity~\cite{Eroshenko:2022nph}. This approach can provide a long--awaited answer to question on how  exactly unitary evolution managing quantum--gravitational degrees of freedom.

%In order to test unitarity one may numerically check if the equality
%\begin{equation}\label{eq:unitarity-test}
%\int{\cal D}c_k^*{\cal D}c_k\;e^{-\int dk\;c_k^*c_k}\langle b|\hat{S}^\dagger|c\rangle\langle c|\hat{S}|a\rangle = \langle b|a\rangle\;
%\end{equation}
%is satisfied. It seems to be completely trivial for any reasonable quantum field theory on flat spacetime. We plan to make examples and counterexamples of such models with unitarity test elsewhere. The gravitational theories require more elaborate numerical approach for which we currently develop a code to calculate semiclassically a number of matrix elements to check if satisfy \eqref{eq:unitarity-test}.

%%%%%%%%%%%%%%%%%%%%%%%%%%%%%%%%
\paragraph{Acknowledgments.} We express gratitude to Dmitry Levkov, Bulat Farkhtdinov, Andrey Shkerin, and Petr Satunin, and Sergey Sibiryakov for interesting discussions. This work was supported by the grant 21-1-4-11-1 of the Foundation for the Advancement of Theoretical Physics and Mathematics ``BASIS''. 

%%%%%%%%%%%%%%%%%%%%%%%%%%%%%%%%
\appendix
%%%%%%%%%%%%%%%%%%%%%%%%%%%%%%%%

\section{Dilaton gravity equations}\label{app:field-eqs}
In this Appendix we derive equations of motion for the linear dilaton gravity described by Eq.~\eqref{eq:grav-action} and interacting with arbitrary matter, and find its solutions. For simplicity we assume no coupling to the dilaton in the matter action, $\delta I_\mathrm{matter}/\delta\phi\equiv 0$. By varying the action $I=I_\mathrm{ldg}+I_\mathrm{matter}$ with respect to the dilaton $\phi$ and the metric $g_{\mu\nu}$ we find field equations,
\begin{gather}
W'(\phi)R=2W''(\phi)\Box\phi+W'''(\phi)((\nabla\phi)^2-\lambda^2)\;, \label{eq:R}\\
-2W'(\phi)\nabla_\mu\nabla_\nu\phi+g_{\mu\nu}\left(2W'(\phi)\Box\phi+W''(\phi)((\nabla\phi)^2-\lambda^2)\right)=T_{\mu\nu}\;, \label{eq:g}
\end{gather}
where $\Box=g^{\mu\nu}\nabla_\mu\nabla_\nu$ is the wave operator and
\begin{equation}
T_{\mu\nu}=-\frac{2}{\sqrt{-g}}\frac{\delta I_\mathrm{matter}}{\delta g^{\mu\nu}}
\end{equation}
is the matter energy--momentum tensor which to be specified.

In order to find a general solution of Eqs.~\eqref{eq:R} and \eqref{eq:g} we choose a diagonal ansatz for the metric,
\begin{equation}\label{eq:nzm}
ds^2=-e^{\nu}dt^2+e^{\zeta}dr^2\;,
\end{equation}
where $\nu=\nu(t,r)$ and $\zeta=\zeta(t,r)$ are arbitrary functions of two coordinates.

One can show the system of Eqs.~\eqref{eq:g} with no matter ($T_{\mu\nu}=0$) is equivalent to
\begin{gather}
\partial_r\nu\,\partial_t\phi+\partial_t\zeta\,\partial_r\phi=2\partial_t\partial_r\phi\;, \label{eq:tr}\\
e^{-\nu}(2\partial_t^2\phi-\partial_t(\nu+\zeta)\partial_t\phi)+e^{-\zeta}(2\partial_r^2\phi-\partial_r(\nu+\zeta)\partial_r\phi)=0\;, \label{eq:ttrr}
\end{gather}
which are ($tr$) and ($tt$)+($rr$) components correspondingly.

The Minkowsi spacetime ($\nu=0=\zeta$) with the dilaton,
\begin{equation}\label{eq:gen-dil}
\phi(t,r)=\pm\lambda r\cdot\cosh\,\theta+\lambda t\cdot\sinh\,\theta+a\;,
\end{equation}
is a solution of Eqs.~\eqref{eq:tr} and \eqref{eq:ttrr}. One can bring Eq.~\eqref{eq:gen-dil} to the ``canonic'' linear dilaton form $\phi=-\lambda r$ by exploiting a Poincare isometry with boost parameter $\theta$, translation $a$, and maybe inversion $r\,\mapsto\,-r$.

Let us impose the coordinate system with $\phi=-\lambda r$. Then it immediately follows from Eqs.~\eqref{eq:tr} and \eqref{eq:ttrr},
\begin{equation}
\zeta=\zeta(r)\;, \qquad\qquad \nu(t,r)=-\zeta(r)+c(t)\;,
\end{equation}
where $c(t)$ is an integration constant which can be always removed by residual time reparametrization invariace in Eq.~\eqref{eq:nzm}.

The system of equations reduces to
\begin{equation}
\partial_r\left((1-e^{-\zeta})W'(-\lambda r)\right)=0\;,
\end{equation}
with a solution $e^{-\zeta(r)}=e^{\nu(r)}=f(r)$ for which Eq.~\eqref{eq:R} is automatically satisfied.

Finally, let us solve Eqs.~\eqref{eq:R}, \eqref{eq:g} with massless scalar field $\chi$,
\begin{equation}
I_\mathrm{matter}=-\frac12\int d^2x\sqrt{-g}(\nabla\chi)^2\;,
\end{equation}
which has the energy--momentum tensor,
\begin{equation}
T_{\mu\nu}=\nabla_\mu\chi\nabla_\nu\chi-\frac12 g_{\mu\nu}(\nabla\chi)^2\;,
\end{equation}
and satisfies the wave equation $\Box\chi=0$.

We introduce the Eddington--Finkelstein metric ansatz,
\begin{equation}
ds^2=-F(v,r)dv^2+2dvdr\;,
\end{equation}
where $v$ is the advanced time. The wave equation reduces to $2\partial_v\chi+F(v,r)\partial_r\chi=0$ which is automatically satisfied for ingoing wave packets $\chi(v,r)\equiv\chi_\mathrm{in}(v)$. The it is easy to show that the only non--vanishing component of the energy--momentum tensor is $T_{vv}=(\partial_v\chi_\mathrm{in})^2$.

By fixing the linear dilaton gauge $\phi=-\lambda r$ one finds the Vaidya soluton,
\begin{equation}
F(v,r)=1+\frac{{\cal M}(v)}{\lambda W'(\phi)}\;,
\end{equation}
where
\begin{equation}
{\cal M}(v)=\int^v_{-\infty}dv'(\partial_v\chi_\mathrm{in}(v'))^2
\end{equation}
is the Bondi mass.

%\section{CFT entropy}\label{app:cft-entropy}
%The interval distance $d_{AB}=\sqrt{(U_A-U_B)(V_B-V_A)}$. 
%one finally obtains
%\begin{equation}\label{eq:cft-entropy}
%S_\mathrm{ent}[\Sigma_{AB}]=\frac{c}{6}\log({d_{AB}}^2e^\rho_A e^\rho_B)\;.
%\end{equation}

\section{QEPs location and entropy}\label{app:islands}
In this Appendix we solve the extremality condition,
\begin{equation}\label{eq:extr-cond-r}
\frac{\partial S_\mathrm{gen}}{\partial r_Q}(r_Q,r_A,M)=0\;,
\end{equation}
where $S_\mathrm{gen}$ is given by Eq.~\eqref{eq:gen-formula-no-time} to determine positions of quantum extremal points $r_Q=r_Q(M,r_A)$.

\begin{figure}[t]
\centerline{
\hspace{0.0cm}\includegraphics[width=8.21cm]{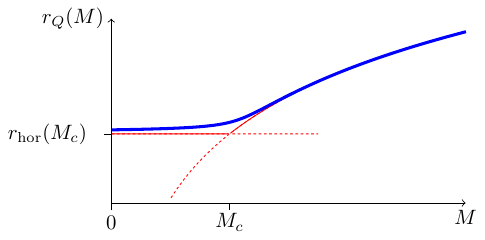}\hspace{1.0cm}\includegraphics[width=7.24cm]{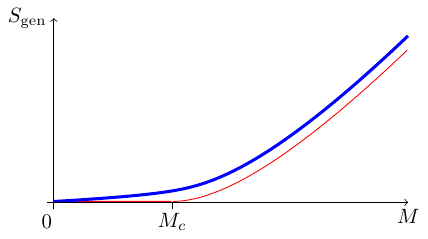}
}
\vspace{2mm} \hspace{4.5cm}(a) \hspace{7.5cm}(b)
\caption{(a) Positions of QEP $r_Q$ at different values of $r_A$ as function of black hole mass for the CGHS model. Smooth blue curve is numerical solution of Eq.~\eqref{eq:tr-cghs} at finite $r_A$ whereas red line corresponds to the limit $r_A\to +\infty$. (b) Generalized entropy functional calculated for finite $r_A$ (thick) and in the limit $r_A\to +\infty$ (thin). } \label{fig:cghs-plots}
\end{figure}

In case of the CGHS model Eq.~\eqref{eq:extr-cond-r} becomes,
\begin{equation}\label{eq:tr-cghs}
\frac1{\lambda}\frac{\partial S_\mathrm{gen}}{\partial r_Q}=16\pi\,e^{2\lambda r_Q}+\frac{2c}{3}\frac1{\sqrt{e^{2\lambda r_Q}-\frac{M}{2\lambda}}-\sqrt{e^{2\lambda r_A}-\frac{M}{2\lambda}}}\frac{e^{2\lambda r_Q}}{\sqrt{e^{2\lambda r_Q}-\frac{M}{2\lambda}}}-\frac{c}{3}=0\;,
\end{equation}
which can be solved numerically, see in Fig.~\ref{fig:cghs-plots}(a).

One can approximately resolve the extremality condition which reduces to
\begin{equation}
\left(e^{2\lambda r_Q}-\frac{c}{48\pi}\right)\sqrt{e^{2\lambda r_Q}-\frac{M}{2\lambda}}=\frac{c}{24\pi}e^{\lambda(2r_Q-r_A)}\;
\end{equation}
in the limit $r_A\gg r_\mathrm{hor}(M)=\log(M/2\lambda)/2\lambda$ so that solution of Eq.~\eqref{eq:tr-cghs} interpolates between two branches,
\begin{equation}\label{eq:branches}
r_Q=\frac1{2\lambda}\log\left(\frac{M}{2\lambda}\right)\;, \;\qquad M>M_c\;\qquad\quad r_Q'=\frac1{2\lambda}\log\left(\frac{c}{48\pi}\right)\;, \qquad M<M_c\;,
\end{equation}
becoming exact in the limit $r_A\to +\infty$, see in Fig.~\ref{fig:cghs-plots}(a). For large black hole mass the QEP lies slightly above the event horizon and for lighter masses gets stuck at critical value of the mass parameter $M_c=c\lambda/24\pi$.

%One finds from perturbative expansion keeping relevant small parameters for each case,
%\begin{equation}
%r_Q(M)\simeq r_\mathrm{hor}(M)+\frac{M e^{-2\lambda r_A}}{\lambda^2\left(\frac{M}{M_c}-1\right)^2}\;,
%\end{equation}
%and
%\begin{equation}
%r_Q(M)\simeq r_\mathrm{hor}(M_c)+\frac{c\,e^{-\lambda r_A}}{48\pi\lambda\sqrt{\frac{c}{48\pi}-\frac{M}{2\lambda}}}\;,
%\end{equation}
%where $r_\mathrm{hor}(M)=\log(M/2\lambda)/2\lambda$ is a position of the event horizon for the CGHS black hole. 

Substituting the infrared regulator,
\begin{equation}
S_0(c,\epsilon,r_A)=-\frac{c}{3}\lambda r_A+\frac{2c}{3}\log(\lambda\epsilon)+\frac{c}{6}\left(\log\left(\frac{c}{48\pi}\right)-1\right)\;,
\end{equation}
and Eq.~\eqref{eq:branches} into Eq.~\eqref{eq:reg-ent} so that $S_\mathrm{ent}(M_c)=0$ one derives Eq.~\eqref{eq:ent-cghs}.

\begin{figure}[t]
\centerline{
\hspace{0.0cm}\includegraphics[width=7.63cm]{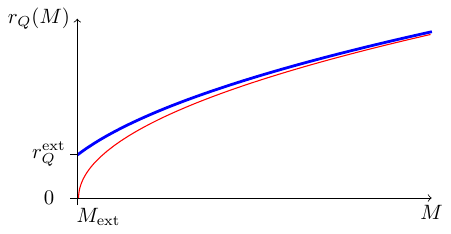}\hspace{1.0cm}\includegraphics[width=7.29cm]{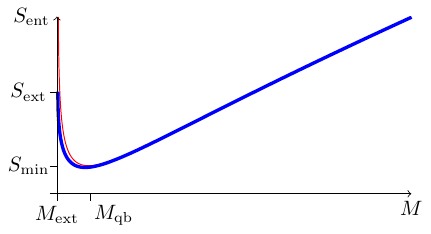}
}
\vspace{2mm} \hspace{4.5cm}(a) \hspace{7.5cm}(b)
\caption{ (a) Position of QEP $r_Q$ at different values of $r_A$ as function of black hole mass for the sinh--CGHS model. Blue curve is numerical solution of extremality condition at finite $r_A$, red line is a solution in the limit $r_A\to +\infty$. (b) Generalized entropy functional calculated for finite $r_A$ (thick) and in the limit $r_A\to +\infty$ (thin). } \label{fig:sinh-cghs-plots}
\end{figure}

In case of the sinh--CGHS model the extremality condition on $r_Q$ can be solved numerically, see in Fig.~\ref{fig:sinh-cghs-plots}(a). One can prove that $r_Q=r_\mathrm{hor}(M)$ from Eq.~\eqref{eq:sh-cghs-hor} in the limit $r_A\to +\infty$ for all masses $M\geq M_\mathrm{ext}$. We can directly check this analytically for the extremal solution $M=M_\mathrm{ext}$. 

By keeping $\lambda r_Q$ one finds from the extremality condition~\eqref{eq:extr-cond-r},
\begin{equation}
1+\frac{M_\mathrm{ext}}{M_c}\lambda r_Q=\frac1{\frac12+\lambda^2r_Q r_A}\;,
\end{equation}
which has an asymptotic solution,
\begin{equation}\label{eq:rq-ext}
\lambda r_Q \simeq \frac1{2\lambda r_A}-\frac1{4\lambda^2{r_A}^2}\frac{M_\mathrm{ext}}{M_c}\;,
\end{equation}
at large $r_A$. Therefore, one concludes from monotonic behaviour in $r_Q(M)$ the QEP always stays close to the event horizon. 

Substitution of Eq.~\eqref{eq:rq-ext} into generalized entropy functional gives a formula,
\begin{equation}\label{eq:div-entr}
S_\mathrm{ext}=\frac{c}{3}\log(2\lambda r_A)+\frac{c}{6}\left(\log\left(\frac{16M_c}{M_\mathrm{ext}}\right)-1\right)+\frac{4\pi M_\mathrm{ext}}{\lambda^2r_A}\;,
\end{equation}
which is divergent in the limit $r_A\to +\infty$. Dependence on the position of the anchor point $A$ in Eq.~\eqref{eq:div-entr} suggests this divergence is related to soft modes living in the near--extremal black hole background.

One can see from Eq.~\eqref{eq:div-entr} the infrared regulator can not be chosen as we did in the previouse subsection by setting the entropy to zero for extremal solutions. What we do instead is to look at large $M$ aymptotics to compare with the CGHS model. One can expect that physics of the Hawking radiation is insensitive to the extremal mass if black holes are comparably large. Therefore, their relative entropies have to match,
\begin{equation}\label{eq:ignorance}
\frac{S_\mathrm{gen}(M,M_\mathrm{ext})}{S_\mathrm{gen}(M,0)}\simeq 1
\end{equation}
as $M\to +\infty$ so that the external observer is ignorant of internal degrees of freedom and can not distinguish large black holes in different models. 

By plugging $r_Q=r_\mathrm{hor}(M)$ into Eq.~\eqref{eq:gen-formula-no-time} and setting $S_0$ so that Eq.~\eqref{eq:ignorance} is satisfied one obtains a nasty crocodile,
\begin{gather}
S_\mathrm{ent}[R]=\frac{4\pi}{\lambda}M\sqrt{1-\left(\frac{M_\mathrm{ext}}{M}\right)^2}+\frac{N}{6}\left[\log\left(\frac{M}{2\lambda}\right)-\log\left(1-\left(\frac{M_\mathrm{ext}}{M}\right)^2\right)\right.\notag\\
\left.-\sqrt{1-\left(\frac{M_\mathrm{ext}}{M}\right)^2}\log\left(\frac{M}{M_\mathrm{ext}}+\sqrt{\left(\frac{M_\mathrm{ext}}{M}\right)^2-1}\right)+\log\left(\frac{1+\frac{M_\mathrm{ext}}{M}-\sqrt{1-\left(\frac{M_\mathrm{ext}}{M}\right)^2}}{1+\frac{M_\mathrm{ext}}{M}+\sqrt{1-\left(\frac{M_\mathrm{ext}}{M}\right)^2}}\right)\right.\notag\\
\left.-1+\log\left(\frac{N}{48\pi}\right)\right]+\frac{N}{3}\left(\sqrt{1-\left(\frac{M_\mathrm{ext}}{M}\right)^2}-1\right)\lambda r_A\;,\label{eq:exact-sh-cghs-gen}
\end{gather}
which has a minimum,
\begin{equation}
S_\mathrm{min}\simeq\frac{c}{6}\left(\log\left(\frac{M_\mathrm{ext}} {M_c}\right)+1-\frac{2M_c}{M_\mathrm{ext}}\right)\;,
\end{equation}
at the value of mass parameter $M_\mathrm{qb}'=M_\mathrm{ext}+2{M_c}^2/M_\mathrm{ext}$, see in Fig.~\ref{fig:sinh-cghs-plots}.

\section{Transition amplitude calculation}\label{app:ampl}
By calculating the total action,
\begin{equation}\label{eq:tot-action}
I_\mathrm{tot}[\Phi]=I_\mathrm{sw}+I_\mathrm{ldg}+I_\mathrm{GH}+I_\mathrm{free}-i\log(\Psi_f^\ast\Psi_i)\;,
\end{equation}
we determine the transition amplitude for the massless particle ${\cal A}_{fi}=\exp(iI_\mathrm{tot}[\Phi])$ in the semiclassical approximation. Let us evaluate all terms from Eq.~\eqref{eq:tot-action} in order of appearance by following guidelines from the technical Appendix from previous paper~\cite{Fitkevich:2020tcj}.

The first term in Eq.~\eqref{eq:tot-action} is the massless particle (shock wave) action,
\begin{equation}\label{eq:null-action}
I_\mathrm{sw}=\int d\zeta\,\Lambda(\zeta)(dx^\mu/d\zeta)^2\;,
\end{equation}
which vanishes on--shell. The equation of motion for the shock wave is $(dx^\mu/d\zeta)^2=0$ which has a general solution $t=\bar{r}(r)+\mathrm{const}$, cf. Eq.~\eqref{eq:tortoise}. By using $t=\bar{r}(r)$ one can express the asymptotic time in the limit $r_{f,i}\to\pm\infty$ as,
\begin{equation}
t_f=r_f+\Delta(M)\;, \qquad\qquad t_i=r_i-\Delta(M)\;,
\end{equation}
where
\begin{equation}\label{eq:delta}
\Delta(M)=\frac{M}{\lambda M_\mathrm{ext}}\arctan\left(\sqrt{\frac{M_\mathrm{ext}+M}{M_\mathrm{ext}-M}}\right)/\sqrt{1-\frac{M^2}{M_\mathrm{ext}^2}}\;,
\end{equation}
for both spacetime patches with mass parameters $M_i$ and $M_f$. We use indices $R$ and $L$ to denote variables related to the right and left spacetime patches correspondingly. Note that the mass parameters are $M_R\equiv M_i$ and $M_L\equiv M_f$ as the shock wave moves from left to right.

The second term is the dilaton gravity bulk action defined by Eq.~\eqref{eq:grav-action} which can be rewritten using field equations as
\begin{equation}\label{eq:grav-ac-onshell}
I_\mathrm{ldg}=\frac{2M_\mathrm{ext}}{\lambda}\int_{\cal M} d^2x\sqrt{-g}\,\frac{\Box\phi}{\cosh(2\phi)}-\frac{2M_\mathrm{ext}}{\lambda}\int_{\partial{\cal M}} ds\cosh(2\phi)\,\kappa\, n^\mu\nabla_\mu\phi\;,
\end{equation}
where $\sigma$ is the proper time/distance on the boundary of spacetime, $n^\mu$ is the outer normal, and $\kappa=n^\mu n_\mu=\pm 1$.

For a static empty spacetime one has $\Box\phi=-\lambda
 f'(r)$ so that Eq.~\eqref{eq:grav-ac-onshell} simplifies,
\begin{multline}
I_\mathrm{ldg}=\int_{t_L,i}^{t_L,f}dt_L\left(\frac{M_f}{\cosh^2(2\lambda r(t_L))}-\frac{M_f}{\cosh^2(2\lambda r_L)}-2M_\mathrm{ext}\cosh(2\lambda r_L)f_L(r_L)\right) \\
+\int_{t_R,i}^{t_R,f}dt_R\left(\frac{M_i}{\cosh^2(2\lambda r_R)}-\frac{M_i}{\cosh^2(2\lambda r(t_R))}+2M_\mathrm{ext}\cosh(2\lambda r_R)f_R(r_R)\right)\;,\label{eq:ldg-simp}
\end{multline}
where $r=r_L$ and $r=r_R$ are the left and right timelike boundaries of spacetime, $n^\mu\partial_\mu\phi=\mp\lambda\sqrt{f}$, and $n^\mu\partial_\mu\phi=0$ for the spacelike boundaries at $t=t_{f,i}$. 

One finds,
\begin{multline}\label{eq:tot-grav-part}
I_\mathrm{ldv}\simeq\int_{-\infty}^{+\infty}dr\left(\frac{M_f}{f_L(r)\cosh^2(2\lambda r)}-\frac{M_i}{f_R(r)\cosh^2(2\lambda r)}\right)\\
=2\lambda M_\mathrm{ext}^2\left(\frac{\Delta(M_f)}{M_f}-\frac{\Delta(M_i)}{M_i}\right)\;,
\end{multline}
where integration variables were changed using the shockwave equations of motion $r=r(t_{R,L})$ and Eq.~\eqref{eq:delta} was applied in the limit $r_{f,i}\to\pm\infty$.

The third contribution in Eq.~\eqref{eq:tot-action} are the Gibbons--Hawking terms,
\begin{equation}
I_\mathrm{GH}=-\frac{M_\mathrm{ext}}{\lambda}\int ds\,\sinh(2\phi)\,\kappa\,(K-K_0)\;,
\end{equation}
where $K=\nabla_\mu n^\mu$ is the extrinsic curvature of the boundary and the parameter $K_0$ removes divergence from Eq.~\eqref{eq:ldg-simp}; one sets $K_0=0$ for spacelike parts of the boundary at $t=t_{f,i}$ and $K_0=\pm 2\lambda$ for timelike parts at $r\to\mp\infty$.

The Gibbons--Hawking terms give non--zero contribution from points $r_\times=r(t_{f,i})$ where the shock wave crosses the Cauchy surfaces $t=t_{f,i}$. Integrating in the vicinity of the crossing point one obtains,
\begin{equation}
2\int ds\,K=\log{\left(\frac{f_L(r_\times)}{f_R(r_\times)}\right)}\;,
\end{equation}
where $K$ is delta--functional.

Contributions from both Cauchy surfaces double each other,
\begin{equation}\label{eq:gibb-haw}
I_\mathrm{GH}=\frac{M_i-M_f}{\lambda}\;,
\end{equation}
in the limit $r_{f,i}\to\pm\infty$.

Notice that expression~\eqref{eq:tot-grav-part} changes its sing if $M_f$ and $M_i$ are swapped. One concludes that total amplitude should suffice
\begin{equation}\label{eq:swap}
\overline{{\cal A}_{fi}(M_f,M_i)}={\cal A}_{fi}(M_i,M_f)\;,
\end{equation}
where overline designates complex conjugation\footnote{$M_f$ and $M_i$ swap change energy sing $E\;\mapsto\; -E$ which gives complex conjugate of amplitude corresponding to evolution backward in time.}.

The last two terms in Eq.~\eqref{eq:tot-action} comes from free evolution operators $\hat{{\cal U}}_0$ and semiclassical wave functions $\Psi_{f,i}=\exp(ip \bar{r}_{f,i})$. The Hamiltonian action of the massless particle with energy $E$ and momentum $p$,
\begin{equation}
I_0=E\int dt-p\int dr\;,
\end{equation}
where integration is along parts of the time contour corresponding to free evolution,
\begin{equation}
I_\mathrm{sw,0}(t_i,0)=p(r_i-\bar{r}_i)-Et_i\;, \qquad\qquad I_\mathrm{sw,0}(0,t_f)=p(\bar{r}_f-r_f)+Et_f\;,
\end{equation}
so that $t=t_R$ is an asymptotic time on the right patch with mass $M_i$ for definiteness. Combining all of this one obtains an expression,
\begin{equation}
I_\mathrm{free}-i\log(\Psi_f^\ast\Psi_i)=(M_i-M_f)(t_f-r_f-t_i+r_i)+iM_f\tau_\mathrm{delay}\;,
\end{equation}
where the last factor comes from gravitational time deficit of a phase factor $e^{-iM_f t}$ from free evolution of the static configuration with mass $M_f$. The delay time $\tau_\mathrm{delay}=2\Delta(M_i)-2\Delta(M_f)$ so that the free action,
\begin{equation}\label{eq:free-part}
I_\mathrm{free}-i\log(\Psi_f^\ast\Psi_i)=2M_i\Delta(M_i)-2M_f\Delta(M_f)\;,
\end{equation}
respects the swap property~\eqref{eq:swap}.

Combining Eqs.~\eqref{eq:tot-grav-part}, \eqref{eq:gibb-haw}, and \eqref{eq:free-part} we find the total action,
\begin{multline}
I_\mathrm{tot}=\frac{M_R-M_L}{\lambda}-\frac{2M_\mathrm{ext}}{\lambda}\sqrt{1-\frac{M_R^2}{M_\mathrm{ext}^2}}\arctan\sqrt{\frac{M_\mathrm{ext}+M_R}{M_\mathrm{ext}-M_R}}\\+\frac{2M_\mathrm{ext}}{\lambda}\sqrt{1-\frac{M_L^2}{M_\mathrm{ext}^2}}\arctan\sqrt{\frac{M_\mathrm{ext}+M_L}{M_\mathrm{ext}-M_L}}\;,
\end{multline}
which encodes both phase and suppression exponent of the semiclassical transition amplitude.

\section{Semiclassical regularization method}\label{app:regul}
In this Appendix we apply a consistent approach for calculating amplitudes of classically forbidden processes~\cite{Bezrukov:2015ufa,Fitkevich:2020tcj}. 

Evidently, the transition amplitude shall not be affected if one insert a unity into the measure ${\cal D}\Phi$ in Eq.~\eqref{eq:path-int},
\begin{equation}
1\equiv\int_0^{+\infty} dT\int_{-i\infty}^{i\infty}\frac{d\varepsilon}{2\pi i}\,e^{\varepsilon(T-T_\mathrm{int}[\Phi])}\;,
\end{equation}
where $T_\mathrm{int}[\Phi]$ is some positive--definite on real solutions and diffeomorphism--invariant functional. It should be finite on any solutions interpolating from asymptotically static configurations with $M_i<M_\mathrm{ext}$ to ones with $M_f<M_\mathrm{ext}$ but diverge as the extremal configuration approaches. Loosely speaking $T_\mathrm{int}[\Phi]$ measures a traversal time for matter passing through the interaction region $|\phi|\lesssim 1$.

One finds a regularized path integral with complex action,
\begin{equation}
I_\varepsilon[\Phi]=I[\Phi]+i\varepsilon\,T_\mathrm{int}[\Phi]-i\varepsilon\,T\;,
\end{equation}
and additional integrations over $\varepsilon$ and $T$. The saddle--point equations,
\begin{equation}\label{eq:regs}
\frac{\delta I[\Phi]}{\delta\Phi}+i\varepsilon\frac{\delta T_\mathrm{int}[\Phi]}{\delta\Phi}=0\;, \qquad T_\mathrm{int}[\Phi]=T\;, \qquad \varepsilon=0\;,
\end{equation}
determine a new solution which should be found by solving first of Eqs.~\eqref{eq:regs} and then taking a limit $\varepsilon\to +0$.

We propose a functional,
\begin{equation}\label{eq:T-f}
T_\mathrm{int}=-\frac{2}{\lambda^2}\int d^2x\sqrt{-g}\,L(\phi)(\lambda^2-(\nabla\phi)^2)^2\;,
\end{equation}
where
\begin{equation}
L(\phi)=\frac{M_\mathrm{ext}^2\cosh^2(2\phi)}{2\lambda^2}(\delta(\phi-\phi_R)-\delta(\phi-\phi_L))
\end{equation}
is localized far from interaction region. For the shock wave moving from left to right one finds,
\begin{equation}
T_\mathrm{int}=2M_i^2\Delta(M_i)-2M_f^2\Delta(M_f)-(M_i^2-M_f^2)(\phi_R-\phi_L)/\lambda\;,
\end{equation}
which satisfies above requirements.

From Eq.~\eqref{eq:g} we derive,
\begin{equation}
e^{-\zeta}\zeta'+2\lambda\tanh(2\lambda r)(1-e^{-\zeta})=-\frac{i\varepsilon\lambda}{2}L(-\lambda r)\tanh(2\lambda r)(1-e^{-\zeta})^2\;,
\end{equation}
where the energy--momentum tensor following from Eq.~\eqref{eq:T-f} was used.

One finds a solution,
\begin{equation}
1-e^{-\zeta}=\frac{M}{M_\mathrm{ext}\cosh(2\lambda r)}\left(1+\frac{i\varepsilon M}{2M_\mathrm{ext}}\int_{-\infty}^{-\lambda r}d\phi\frac{L(\phi)}{\cosh^2(2\phi)}\right)^{-1}\;.
\end{equation}
which reads that mass acquires a positive imaginary shift,
\begin{equation}
M_{i,f}\;\mapsto\;M_{i,f}\left(1-\frac{i\varepsilon M_\mathrm{ext}M_{i,f}}{4\lambda^2}\right)^{-1}\simeq M_{i,f}+i\varepsilon'\;, \qquad \varepsilon,\,\varepsilon'>0\;,
\end{equation}
validating the analytic continuation in Eq.~\eqref{eq:anal-cont}.

%\section{Global structure of eternal black hole in sinh-CGHS model}
%Now let us turn to the regular black holes. Let us consider a following form of the eternal black hole metric,
%\begin{equation}\label{eq:extended-metric}
%ds^2=\frac{f(r)}{4\pi^2{T_H}^2g(r)}(-dT^2+dR^2)\;,
%\end{equation}
%where the function $f(r)$ is defined in Sec~\ref{sec:models} and the function
%\begin{equation}
%g(r)=\frac{\left(1+\frac{M_\mathrm{ext}}{M}\right)\tanh(\lambda r)-\frac{2\pi T_H}{\lambda}}{\left(1+\frac{M_\mathrm{ext}}{M}\right)\tanh(\lambda r)+\frac{2\pi T_H}{\lambda}}\exp(4\pi T_H r)\;
%\end{equation}
%satisfies $f(r)g'(r)=4\pi T_H g(r)$. Coordinates $(t,r)$ and $(T,R)$ are related by a transformation
%\begin{gather}
%T=\sqrt{g(r)}\sinh(2\pi T_H t)\;, \qquad R=\sqrt{g(r)}\cosh(2\pi T_H t)\;, \quad \mathrm{at}\;\; |R|>|T|\;, \\
%T=\sqrt{-g(r)}\cosh(2\pi T_H t)\;, \qquad R=\sqrt{-g(r)}\sinh(2\pi T_H t)\; \quad \mathrm{at}\;\; |R|<|T|\;,
%\end{gather}
%which is direct analogue of the Kruskal-Szekeres extension.
%
%Notice that the metric \eqref{eq:extended-metric} has imaginary time period equal to the inverse Hawking temperature $1/T_H$. Therefore, the vacuum defined with respect to the coordinates $(T,R)$ is the Hartle-Hawking vacuum, and it is an appropriate frame to evaluate the generalised entropy.

%%%%%%%%%%%%%%%%%%%%%%%%%%%%%%%%%%%%%%%%%

\end{document}